\documentclass[twocolumn,prb,showpacs,amsmath,amssymb,floatfix]{revtex4}

\usepackage{graphicx}
\usepackage{dcolumn}
\usepackage{bm}

\begin{document}

\title{Dark-bright magneto-exciton mixing induced by Coulomb
interaction\\ in strained quantum wells}

\author{Y. D. Jho,$^{1,3}$ F. V. Kyrychenko,$^{1}$ J.
Kono,$^{2}$ X. Wei,$^{3}$ S. A. Crooker,$^{4}$ G. D.
Sanders,$^{1}$ D. H. Reitze,$^{1}$ C. J. Stanton,$^{1}$ and G. S.
Solomon$^{5}$}
\affiliation{$^1$Department of Physics, University of Florida,
Gainesville, FL 32611\\ $^2$Department of Electrical and
Computer Engineering, Rice University, Houston, TX 77005\\
$^3$National High Magnetic Field Laboratory, Florida State
University, Tallahassee, FL 32310\\ $^4$National High Magnetic
Field Laboratory, Los Alamos National Laboratory, Los Alamos, NM
87545\\ $^5$Solid-State Laboratories, Stanford University,
Stanford, CA 94305}

\begin{abstract}
Coupled magneto-exciton states between allowed (`bright') and
forbidden (`dark') transitions are found in absorption spectra of
strained In$_{0.2}$Ga$_{0.8}$As/GaAs quantum wells with increasing
magnetic field up to 30 T. We find large ($\sim$ 10 meV) energy
splittings in the mixed states.  The observed anticrossing
behavior is independent of polarization, and sensitive only to the
parity of the quantum confined states. Detailed experimental and
theoretical investigations indicate that the excitonic Coulomb
interaction rather than valence band complexity is responsible for
the splittings.  In addition, we determine the spin composition of
the mixed states.
\end{abstract}
 \maketitle

When energy levels of excited states are tuned to the same energy,
one can observe either crossing or anti-crossing behavior,
depending on the coupling character of the external
perturbations.\cite{Eck} If the coupling matrix element of
perturbing terms is non-vanishing, their corresponding
wavefunctions are mixed so that the crossing is suppressed and
replaced by anti-crossing behavior. In semiconductor
heterostructures, valence-band complexity arising from spin-orbit
coupling has been extensively investigated in GaAs-based quantum
wells (QWs) and proposed as the main coupling mechanism.
\cite{Vina1,Bauer,Magri,Miller,Molenkamp,Sanders} Due to the
relatively small heavy-hole-light-hole (HH-LH) separation and
resulting close proximity of valence band manifolds, the coupling
of HH-LH exciton states in GaAs QWs can be substantiated through
strong modifications of in-plane effective
masses\cite{Vina2,Sanders2} and transitions that are nominally
forbidden in a simple two-band model can be observed
optically.\cite{Magri,Miller,Molenkamp,Theis} While a number of
studies on valence-band complexity have been reported for
exciton-mixing in QW's, it is conceivable that other mechanisms
can also determine the nature of the interaction. Recently,
anti-crossing involving bright and dark excitons in single quantum
dot molecules due to coherent coupling between the states of the
two dots has been observed. \cite{Ortner}  In addition, large
splittings of cyclotron resonance lines in Al$_x$Ga$_{1-x}$N/GaN
heterostructures have been reported \cite{Syed}, indicating that
the origins of anti-crossing interactions in high magnetic fields
are not completely understood.

As a way to investigate whether exciton mixing is significantly
influenced by other mechanisms, indium can be incorporated into
the QW layer to induce strain. Since the lattice constant in
In$_x$Ga$_{1-x}$As QW is larger than that of surrounding GaAs
layers, a biaxial compressive strain is introduced. The strain
lowers the LH energy relative to the HH energy,\cite{Huang}
thereby reducing the hole-level coupling and its corresponding
repulsive interaction between LH and HH subbands.\cite{Sanders2}
Thus, one can minimize the effects of valence band complexity in
In$_x$Ga$_{1-x}$As QWs and obtain discrete optical spectra of HH
and LH excitons in the high magnetic field regime. In addition,
QWs provide lighter in-plane masses of HHs than that of
LHs,\cite{Sanders,Miller2,Greene} thus magneto-exciton sublevels
evolve differently in applied magnetic
fields.\cite{Bauer,Vina2,Yang} Thus, In$_x$Ga$_{1-x}$As/GaAs QWs
are ideally suited for exploring the nature of the mixing
interaction.

\begin{figure}[tbp]
\includegraphics[scale=0.5,trim=30 0 0 0] {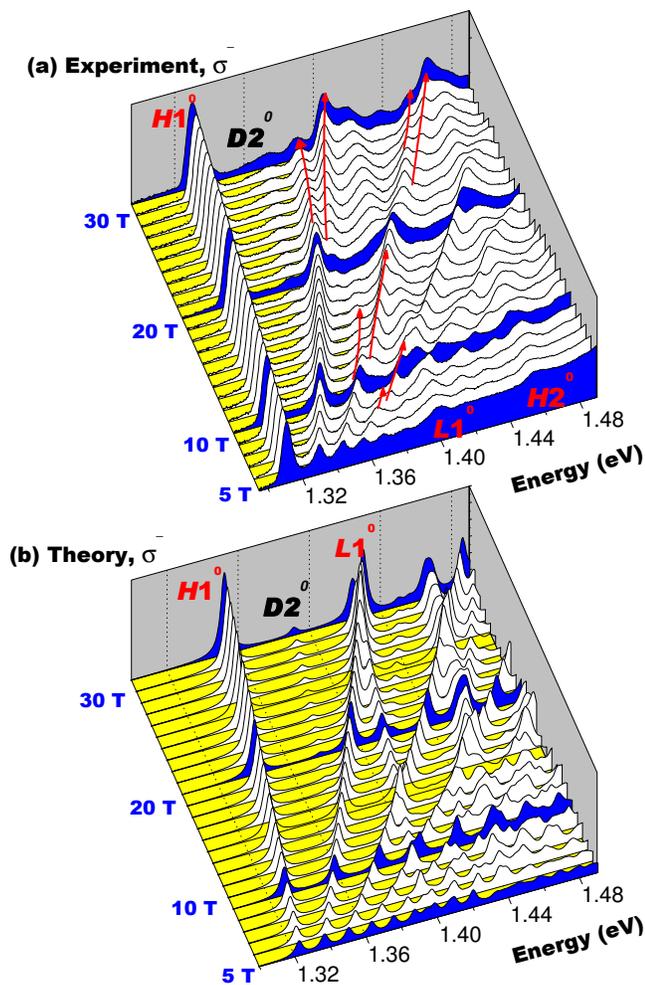}
\caption{(a) Experimental magneto-absorption spectra for
In$_{0.2}$Ga$_{0.8}$As/GaAs multiple quantum wells with $\sigma^-$
polarization at 4.2 K for various fields up to 30 T. At zero
field, three prominent peaks, indicated by $H1^0$, $L1^0$, and
$H2^0$, are lowest HH exciton, lowest LH exciton, and second
lowest HH exciton state, respectively. The lines $D2^0$ and $D2^1$
are assigned to the lowest and second lowest ``dark''
magneto-excitons asociated with the $n_e$ = 1 electrons and $n_h$
= 2 heavy holes. The arrows are guides to the eye on separated
magneto-exciton peaks. (b) Theoretical absorption spectra computed
using an 8 band Pidgeon-Brown effective mass model. The dark state
$D2^0$ is clearly visible in the spectra above 20 T. \label{fig1}}
\end{figure}

We report the observation of coupled states between forbidden and
allowed interband transitions in strained In$_x$Ga$_{1-x}$As QW
structures in strong magnetic fields. A new kind of mixed
magneto-exciton state is found in a system where the anti-crossing
between hole states itself is suppressed.  The samples, consisting
of 15 layers of 8 nm thick QWs separated by 15 nm GaAs barriers,
were grown by molecular beam epitaxy at 390 $^{\circ}$C. To
investigate the character of mixed states at low densities,
absorption measurements were carried out in
In$_{0.2}$Ga$_{0.8}$As/GaAs multiple quantum wells for
$\sigma^-$ and $\sigma^+$ polarizations at 4.2 K. Magnetic
fields perpendicular to the QW plane (Faraday geometry) were
applied up to 30 T (45 T) using a resistive Bitter-type magnet
(hybrid resistive-superconducting magnet) at 4.2 K. White-light
from a tungsten-lamp was used as the excitation source. Both
excitation and collection were performed through an optical
fiber at normal incidence to the sample surface. In order to
investigate exciton mixing at higher carrier densities, we also
performed photoluminescence (PL) on our sample as a function of
excitation power.  For these experiments, a linearly polarized
150 fs, 775 nm pulse from a chirped pulse amplifier (Clark-MXR
CPA-2001) was focused in free space to a 500 $\mu$m spot on the
sample. Unpolarized PL was collected using an fiberoptic probe
from the backside of the sample and examined as a function of
both magnetic field and excitation power.

In order to specify a quantum well exciton state in a magnetic
field, we use the high-field Landau notation (as opposed to the
low-field excitonic notation); for low-field---high-field
correspondence, see, e.g., ref. \cite{MacDonald}.  Each state has
four indices, i.e., $N$, $M$, $n_e$, and $n_h$ (or $n_l$).  Here
$N$ and $M$ are electron and hole Landau quantum numbers,
respectively ($N,M$ = 0, 1, 2, ...).  The electron and HH (LH) QW
energy levels are denoted by $n_e$ and $n_h$ ($n_l$).  For the
present work, the most relevant QW level indices are $n_e$ = 1 and
2, $n_h$ = 1, 2, and 3, and $n_l$ = 1.  For convenience, we are
neglecting the spin index and center-of-mass momentum of each
magneto-exciton state. In an ideal quantum well, interband
transitions occur only when $N$ = $M$ and $n_e$ = $n_h$ ($n_l$).
However, $n_e \neq n_h(n_l)$ transitions are usually weakly
allowed in real quantum wells due to perturbations such as strain,
potential asymmetry, or valence-band mixing.  We denote bright
(i.e., optically active) HH (LH) exciton states
$|N,M,n_e,n_h(n_l)\rangle$ = $|N,N,n,n\rangle$ as $Hn^N$ $(Ln^N)$
and dark (i.e., optically inactive) HH exciton states
$|N,M,n_e,n_h\rangle$ = $|N,N,1,n_h(\neq 1) \rangle $ as $Dn^N$.
Thus, all the dark states we are considering here are associated
with $n_e$ = 1 electrons; note also that the Landau quantum number
is always conserved ($N$ = $M$).

Figure 1(a) displays a series of representative absorption spectra
as a function of magnetic field for the case of $\sigma^{-}$
polarization.
Note that at zero field, the HH and LH splitting is large
($\sim$100 meV) due to strain. As the field is increased, four
effects are observed: (1) the absorption spectrum evolves from a
step-like two-dimensional to a delta-function-like
zero-dimensional density of states; (2) magneto-exciton levels are
resolved up to $N$ = 6 ($N$ = 2) for the HH (LH) excitons; (3) the
dark state $D2^N$ develops into a clearly resolved distinct peak;
and (4) significantly, the $H1^N$ subbands with $N \geq 1$ split
into two lines (indicated by the arrows in Fig. 1(a)). The energy
at which the splitting occurs is found to be roughly the same
distance below $L1^0$ for each state and is independent of
polarization. The observation of the normally parity-forbidden
$D2^0$ transition arises from the broken inversion symmetry in the
presence of the magnetic field.

For comparison, Figure 1(b) shows results of a corresponding
theoretical simulation based on an 8 band Pidgeon-Brown model
\cite{pidgeon,sanders2003} of an undoped
In$_{0.2}$Ga$_{0.8}$As/GaAs superlattice on a GaAs substrate.
Superlattice effects are obtained by finite differencing the 8
band Hamiltonian.  While this model explicitly incorporates
pseudomorphic strain in superlattice bandstructure, excitonic
effects resulting from the Coulomb interaction are not included.
In these calculations, the conduction band offset is taken to be
0.7 of the total band offset and the strain in the wells is
determined by the average lattice constant in the superlattice
while the barrier lattice constant is pinned to the substrate
lattice constant \cite{munekata}.  To directly compare the
experiments with a theory without Coulomb interaction, we display
Fig. 1 only over the field range higher than 5 T. This theory
reproduces the experimental features reasonably well, with one
exception, namely that {\em there is no evidence of anti-crossing
of the $H1^N-D3^0$ and $H1^N-D3^1$ states}.

\begin{figure}
\rotatebox{-90}{\includegraphics[scale=0.33,trim=50 50 50 0]
{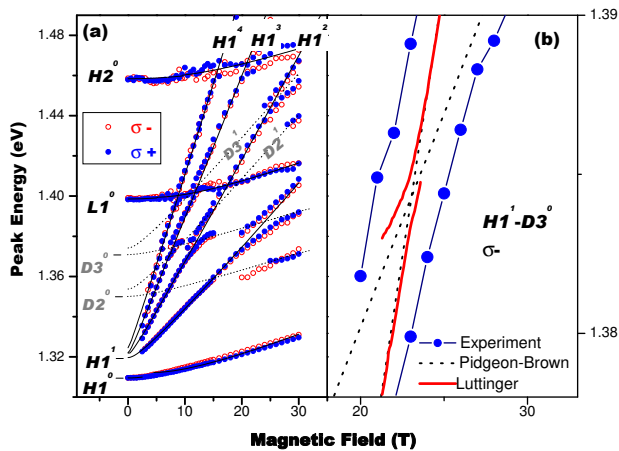}} \caption{(a) Magneto-exciton spectral line positions
obtained by fitting absorption data for $ \sigma^+$ (solid
circles) and $\sigma^-$ (hollow circles) from 0 to 30 T. The solid
(dotted) lines depict the fitting curves of 2D magnet-exciton
transitions for nominally allowed (forbidden) states. The $H1^5$,
$H1^6$, $L1^1$, and $L1^2$ transitions are omitted for clarity.
(b) A comparison of the experimental absorption spectrum for
$\sigma^-$ with theory at the $H1^1 - D3^0$ crossing point. The
dotted line is computed from an 8 band Pidgeon-Brown effective
mass theory without Coulomb interaction. The red line is
calculated using a Luttinger Hamiltonian incorporating the Coulomb
interaction. \label{fig2}}
\end{figure}

Figure 2 (a) shows the prominent exciton lines obtained from
fitting the experimental data in Figure 1 (a).  The filled
(hollow) circles are obtained from measurements with $\sigma^+$
($\sigma^-$ ) polarizations. $H1^5$, $H1^6$, $L1^1$, and $L1^2$
are omitted for clarity although they were experimentally
observed. In order to better visualize the data, the solid lines
are simple parametric curves obtained using the reduced effective
mass $\mu$, the exciton binding energy $E_X$, and the QW energy
gap as fitting parameters (neglecting spin
splitting).\cite{Akimoto} Two features are noteworthy. First, our
data shows that $H_3$ is well separated and not coupled to $L_1$.
Thus, it should not acquire `bright' character through
valence-band mixing.\cite{Miller,Theis} In addition, the Landau
subbands of $H1^N$ are split into two lines below (above) $L1^0$
exactly around the $D3^0$ ($D3^1$) energy position, while the more
optically active $D2^0$ and $D2^1$ do not affect the $H1^N$
subbands. Therefore, the splitting is sensitive to the parity of
the dark and bright exciton branches. Figure 2 (b) displays the
$H1^1 - D3^0$ crossing point, comparing the experimental
absorption data with theoretical predictions. Experimentally, we
find a large ($\sim$ 9 meV) splitting in the energy of the states
near 25 T. However, the Pidgeon-Brown model (dotted line) predicts
no splitting {\em even though it incorporates full valence band
complexity}.  This is true for all $H1^N$-$D3^{N'}$ level
crossings as well.

The failure of a model incorporating valence band complexity to
explain these anti-crossings leads us to consider other
mechanisms. There are two lines of evidence which suggest that the
excitonic Coulomb interaction plays an important role in
determining the splitting energies at the mixing points. First, we
found that at high carrier densities the anti-crossing disappears.
Figure 3 displays field-dependent magneto-photoluminescence (MPL)
spectra upon excitation by 775 nm, 150 fs, 20 GW/cm$^{-2}$ pulses
from a high power chirped pulse amplifier (CPA).  At these
excitation powers, carrier density of $7.5\times 10^{12}$
cm$^{-2}$ are generated. The MPL spectra show various higher
Landau-level states, but we do not observe any anti-crossing even
though clearly resolved higher LL states are visible.  This is
consistent with the mixing having a Coulombic origin. At high
densities, the Coulomb interaction is screened out, and excitonic
states are replaced with magneto-plasmonic behavior.\cite{Butov}

Second, when the excitonic Coulomb interaction is explicitly
included in our calculations, we do observe anti-crossing
behavior.  We developed a comprehensive theory of magnetoexcitons
in quantum wells. The theory is similar to that of Bauer and Ando
\cite{Bauer}, but modified to be valid for high magnetic fields.
For holes, our theory takes into account the Luttinger Hamiltonian
in the axial approximation. The details of the calculations will
be published elsewhere.  The thick line in Figure 2 (b) presents
the results of the Luttinger Hamiltonian for this system including
the Coulomb interaction between electrons and holes. Our
calculations show that Coulomb interaction do results in
anticrossing of $H1^1$ and $D3^0$ states (Fig. 2b), however its
magnitude (2 meV) is considerably smaller than the measured value
of 9 meV.

\begin{figure}
\rotatebox{-90}{\includegraphics[scale=0.35,trim=50 50 50 0]
{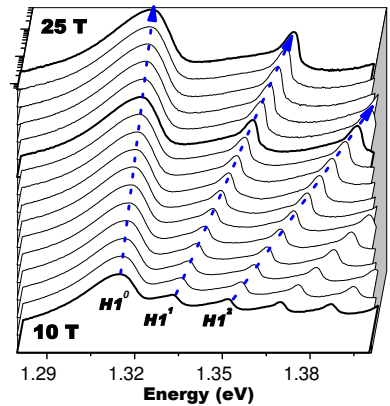}} \caption{Magneto-photoluminescence spectra obtained
after excitation by high power femtosecond pulses.  In contrast to
the CW absorption spectra, no anti-crossing behavior is seen.
\label{fig3}}
\end{figure}

While the Coulomb interaction is important in originating the
anti-crossing, the large difference between the measured and
theoretical energy splitting indicates that other mechanisms are
also involved. We believe that the presence of uniaxial strain in
the plane of the quantum well in our highly strained samples might
play a crucial role. This strain breaks the rotational symmetry of
the problem (which was used in our calculations) and this may
substantially increase the magnitude of anti-crossing matrix
elements

In addition to the optical character of the mixed states, our
experiments also allow us to investigate how the mixed states
`share' their spin (g-factor) character. Near the anti-crossing
point, the wavefunction $\psi_{H1^N}$ mixes with $\psi_{D3^{N'}}$
to reveal a new set of wavefunctions:
\begin{eqnarray}
\psi_1 \, & = \, C_1 \psi_{D3^{N'}} \, + \, C_2 \psi_{H1^N} \\
\psi_2 \, & = \, C_2 \psi_{D3^{N'}} \, - \, C_1 \psi_{H1^N} \, \,
,
\end{eqnarray}
where coupling coefficients $C_1$ and $C_2$, which are the
functions of the Coulomb interaction potential $V_{N,N'}$, are
normalized to unity ($C_1^2+C_2^2=1$). Since $\psi_{D3^{N'}}$ is
optically inactive, the optical character of the new states
$\psi_{1,2}$ preserves the identity of $\psi_{H1^N}$ by sharing
its original oscillator strength. Hence, they are polarized in the
same way. We can describe their polarization state by examining
the Zeeman-splitting $\delta = E(\sigma^+) - E(\sigma^-) = -g_{ex}
\mu_0 B$ between $\sigma^+$ and $\sigma^+$ polarizations, where
$g_{ex}$ is the exciton g-factor and $\mu_0$ is the Bohr magneton.
$g_{ex}$ is the sum of effective hole g-factor $g_h$, which
consists of valence-band Luttinger parameters \cite{Snelling}, and
electron g-factor $g_e$.

Figure 4(a) shows measured $\delta$ for $H1^N$ transitions with
$N$ up to 3. The detailed shifts of $\delta$ for various coupled
peaks are shown in Fig. 4(b) for 12 T and 4(c) 23 T.  From Fig.
4(a), in all cases $g_{ex}$ initially is negative and decreasing
in value at low fields, reaching a minimum value and then
increasing. Within a field range of 45 T, $H1^N$ levels with
$N\geq 1$ show a sign reversal of $g_{ex}$. The field at which the
sign change occurs scales roughly inversely with sub-band index
$N$.  That the spin character of the dark states arises from their
bright companion can be seen by comparing the $H1^N$, $D3^{N'}$,
and $L1^0$ peaks in Figs. 4(b) and (c).  For the three mixed
branches (1) $H1^2-D3^0$ [in Fig. 4(b)], (2) $H1^1-D3^0$ , and (3)
$H1^2-D3^1$ [in Fig. 4(c)], the sign of $\delta$ ($g_{ex}$) for
mixed states follow $H1^N$, regardless of the subband index ($N$,
$N'$). In contrast, the light holes $L1^0$ shift in the opposite
direction, and thus possessing the opposite sign of $g_{ex}$ in
both cases. This further indicates that the mixing is obtained
without interference from the LH excitons.

\begin{figure}
\rotatebox{-90}{\includegraphics[scale=0.33,trim=40 50 50 0]
{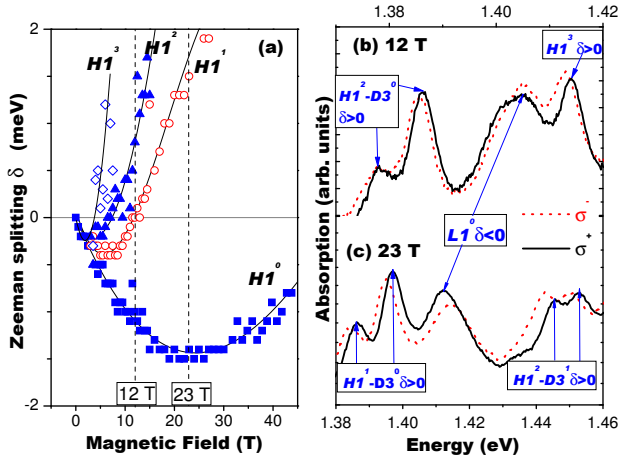}} \caption{(a) Field dependent Zeeman splitting for the
$H1^N$ subbands. The solid lines are a guide to eye. At the fields
denoted by dashed vertical line in (a), the individual $\sigma^+$
(solid lines) and $\sigma^-$ (dotted lines) spectra are plotted
for 12 T (b) and 23 T (c).\label{fig4}}
\end{figure}

In conclusion, we have presented a systematic study of mixed
states of dark and bright magneto-excitons in strong magnetic
field, wherein clear parity dependence was demonstrated. The
polarization and parity dependence are understood through a
Coulomb-interaction-mediated coupling between the same parity
states as well as strain. Qualitative agreements were achieved on
mixing strength and polarization dependence of mixed state.
Field-dependent Zeeman splitting suggests that the spin character
of the dark states comes from their bright exciton partners though
the coupled states are linear combination of dark and bright
excitons whose energy variations are redefined through
off-diagonal matrix terms in the Luttinger Hamiltonian.

\begin{acknowledgements}
This work was supported by DARPA through Grant No.
MDA972-00-1-0034, the NSF ITR program through grant DMR-032547,
and the NHMFL In-house Science Program.
\end{acknowledgements}


\begin{references}
\bibitem{Eck} T. G. Eck, L. L. Foldy, and H. Wieder, Phys. Rev.
Lett. \textbf{10}, 239 (1963).
\bibitem{Vina1} L. Vi\~{n}a \textit{et al.}, Phys. Rev. Lett. \textbf{58}, 832
(1987).
\bibitem{Bauer} G. E. W. Bauer and T. Ando, Phys. Rev. B \textbf{38}, 6015 (1988), and references therein.
\bibitem{Magri} R. Magri and A. Zunger, Phys. Rev. B, \textbf{62},
10364 (2000).
\bibitem{Miller} R. C. Miller \textit{et al.}, Phys. Rev. B \textbf{32}, 8452
(1985).
\bibitem{Molenkamp} L. W. Molenkamp \textit{et al.}, Phys. Rev. B
\textbf{38}, 6147 (1988).
\bibitem{Sanders} G.D. Sanders and Y.C. Chang, Phys. Rev. B
\textbf{32}, 5517 (1985); \textbf{35}, 1300 (1987).
\bibitem{Vina2} L. Vi\~{n}a \textit{et al.}, Phys. Rev. B \textbf{47}, 13926
(1993).
\bibitem{Sanders2} G. D. Sanders \textit{et al.}, Phys. Rev. B
\textbf{50}, 8539 (1994).
\bibitem{Syed} S. Syed, M. J. Manfra, Y. J. Wang, H. L. Stormer, and R. J.
Molnar, Phys. Rev. B \textbf{67}, 241304 (2003).
\bibitem{Theis} W.M. Theis \textit{et
al.}, Phys. Rev. B \textbf{39}, R1442 (1989).
\bibitem{Ortner} G. Ortner, M. Bayer, A. Larionow, V.B. Timofeev, A. Forchel, Y.B. Lyanda-Geller, T.L. Reinecke, P. Hawrylak, S. Fafard, and Z. Wasilewski, Phys. Rev. Lett. \textbf{90}, 86404 (2003).
\bibitem{Huang} K. F. Huang, K. Tai, S. N. G. Chu, and A. Y. Cho,
Appl. Phys. lett. \textbf{54}, 2026 (1989).
\bibitem{Miller2} R.C. Miller, D.A. Keinman, W.T. Tsang, and A.C.
Gossard, Phys. Rev. B \textbf{24}, 1134 (1981).
\bibitem{Greene} R.L. Greene, K.K. Bajaj, and E.E. Phelps, Phys.
Rev. B \textbf{29}, 1807 (1984).
\bibitem{Yang} S.-R. Eric Yang and L.J. Sham, Phys. Rev. Lett.
\textbf{58}, 2598 (1987).
\bibitem{pidgeon} C. K. Pidgeon and R. N. Brown, Phys. Rev. \textbf{146}, 575 (1966).
\bibitem{sanders2003} G. D. Sanders \textit{et al.}, Phys. Rev. B \textbf{68}, 165202 (2003).
\bibitem{Cheung} S.K. Cheung \textit{et al.}, J. Appl. Phys.
\textbf{81}, 497 (1997).
\bibitem{Akimoto} O. Akimoto and H. Hasegawa, J. Phys. Soc. Jpn,
\textbf{22}, 181 (1967).
\bibitem{MacDonald} A. H. MacDonald and D. S. Ritchie, Phys. Rev.
B \textbf{33}, 8336 (1986).
\bibitem{Bastard} G. Bastard, in \textit{Wave mechanics applied to semiconductor heterostructures},
Paris: \'{E}ditions de Physique, p 317.
\bibitem{Butov} L.V. Butov, V.D. Egorov, and V.D. Kulakovskii,
Phys. Rev. B \textbf{46}, 15156 (1992).
\bibitem{Snelling} M. J. Snelling \textit{et al.}, Phys. Rev B \textbf{45}, 3922 (1992).
\bibitem{munekata} H. Munekata, personal communication.
\end{references}
\end{document}